\documentclass[11pt]{article}
\usepackage{epsfig}

\newcommand{\be}{\begin{equation}}
\newcommand{\ee}{\end{equation}}
\newcommand{\bea}{\begin{eqnarray}}
\newcommand{\eea}{\end{eqnarray}}
\newcommand{\sn}{{\rm sn}}
\newcommand{\cn}{{\rm cn}}
\newcommand{\dn}{{\rm dn}}
\newcommand{\sech}{{\rm sech}}

\begin{document}
\vspace{.5in}
\begin{center}
{\LARGE \bf Supersymmetry in Quantum Mechanics}
\end{center}
\vspace{.5in}
\begin{center}
{\Large \mbox{AVINASH KHARE}}
\end{center}
\vspace{.1in}
\begin{center}
{\Large \bf Institute of
Physics, Sachivalaya Marg, Bhubaneswar 751005, Orissa, India}  
\end{center}
\vspace{1.4in}

\begin{abstract}

An elementary introduction is given to the subject of Supersymmetry in
Quantum Mechanics which can be understood and appreciated by any one
who has taken a first course in quantum mechanics. We demonstrate with
explicit examples that given a solvable problem in quantum mechanics 
with n bound states, one can construct new exactly solvable 
n Hamiltonians having
n-1,n-2,...,0 bound states. The relationship between the eigenvalues,
eigenfunctions and scattering matrix of the supersymmetric partner
potentials is derived and a class of reflectionless potentials are
explicitly constructed. We extend the operator method of solving the
one-dimensional harmonic oscillator problem to a class of potentials
called shape invariant potentials. It is worth emphasizing that this
class includes almost all the solvable problems that are found in the
standard text books on quantum mechanics. Further, we show that given
any potential with at least one bound state, one can very easily
construct one continuous parameter family of potentials having same
eigenvalues and s-matrix. The supersymmetry inspired WKB approximation
(SWKB) is also discussed and it is shown that unlike the usual WKB, the
lowest order SWKB approximation is exact for the shape invariant
potentials and further, this approximation is not
only exact for large quantum numbers but by construction, it is also
exact for the ground state. Finally, we also construct new exactly
solvable 
periodic potentials by using the machinery of supersymmetric quantum
mechanics.
\end{abstract}

\section{Introduction}

Supersymmetry  (SUSY) is a symmetry between fermions and bosons. It
was first introduced in High Energy Physics in an attempt 
to obtain a unified description of all basic interactions of 
nature. It is a highly unusual symmetry since  
fermions and bosons have very different properties. For
example, while identical bosons condense, in view of Pauli exclusion
principle, no two identical fermions
can occupy the same state! Thus it is quite remarkable that one could
implement such a symmetry. 
 The algebra involved in  SUSY is a graded Lie
algebra which closes under a combination of commutation and 
anti-commutation relations. In the context of particle physics, SUSY
predicts that corresponding to every basic constituent of nature,
there should be a supersymmetric partner with spin
differing by half-integral unit. Further it predicts that the two
supersymmetric partners must have identical mass in case supersymmetry
is a good symmetry of nature. In the context of unified theory for the
basic interactions of nature, supersymmetry predicts the existence of  
SUSY partners of all the basic constituents of nature, i.e.  
SUSY partners of 6 quarks, 6 leptons 
and the corresponding gauge quanta (8 gluons, photon,
$Z^{0}, W^{\pm}$). 
The fact that no scalar electron 
has been experimentally observed with mass less than about 100 GeV
(while the electron mass is only 0.5 MeV) means that SUSY must be a
badly broken symmetry of nature. Once this realization came, people
started to understand the difficult question of spontaneous SUSY breaking in
quantum field theories. It is in this context that Witten\cite{wit81} suggested in
1981 that perhaps one should first understand the question 
of SUSY breaking in the
simpler setting of nonrelativistic quantum mechanics and this is how 
the area of SUSY quantum mechanics was born.

   Once people started studying various aspects of supersymmetric quantum 
mechanics
(SQM), it was soon clear that this field was interesting in its 
own right, not just
as a model for testing concepts of SUSY field theories. In the last 20
years, SQM 
has given us deep insight into several aspects of standard
nonrelativistic quantum mechanics. The purpose of these lectures is to
give a brief introduction to some of these ideas. For example,

\begin{enumerate}

\item It is well known that the infinite square well is one of the
simplest exactly solvable problem in nonrelativistic QM and the energy
eigenvalues are given by $E_n = c (n+1)^2$ with $c$ being a constant.
Are there other potentials for which the energy eigenvalues have a
similar form and is there a simple way of obtaining these potentials.

\item Free particle is obviously the simplest (and in a way trivial) 
example in QM with no
bound states, no reflection, and the transmission probability being
unity. Are there some nontrivial potentials for which also there is
no reflection and is it possible to easily construct them?

\item Among the large number of possible potentials, only a very few  
are analytically solvable. What is so special about these ``solvable''
potentials? 

\item One problem which all of us solve by two different methods (i.e.
by directly solving Schr\"odinger equation and by operator method) is
the one dimensional harmonic oscillator potential. Can one extend this
operator method to a class of potentials? In this context, it is worth
recalling that the operator method of solving the one dimensional
harmonic oscillator problem is very fundamental and in fact forms the basis 
of quantum
field theory as well as many body theory.

\item Given a potential $V(x)$, the corresponding energy eigenvalues
$E_n$, and the scattering matrix (i.e. the reflection and transmission
coefficients $R(k),T(k)$ in the one dimensional case or the phase
shifts in the three dimensional case) are unique. Is the converse also
true, i.e. given all the energy eigenvalues and $R(k)$ and $T(k)$ at all
energies, is the corresponding potential unique? If not, then how does
one construct these strictly isospectral potentials having same
$E_n,R(k),T(k)$? 

\item A related question is about the construction of the soliton
solutions of the KdV and other nonlinear equations. Can these be
easily constructed from the formalism of SQM?

\item WKB is one of the celebrated semiclassical approximation scheme
which is expected to be exact for large quantum numbers but usually
is not as good for small quantum numbers. Is it possible to have a
modified scheme which by construction would not only be exact for
large quantum numbers but even for the ground state so that 
it has a chance to do better even when the quantum numbers are neither
too large nor too small? Further, 
the lowest order WKB is exact in the case of only
two potentials, i.e. the one dimensional oscillator and the Morse
potentials. Can one construct a modified WKB scheme for which the
lowest order approximation will be exact for a class of potentials? 

\end{enumerate}

The purpose of these lectures is to answer some of the questions
raised above. 
We shall first discuss the basic formalism 
of SQM and then discuss some of these issues. For pedagogical
reason, we have kept these lectures at an elementary level. More
details as well as discussion about several other topics can be
obtained from our book \cite{cks01} and Physics Reports
\cite{cks95} on this subject.
A clarification is in order here. Unlike in SUSY quantum
field theory, in SQM, supersymmetric partners are not fermions and
bosons, instead here, SUSY relates the eigenstates and S-matrix of
the two partner Hamiltonians. 

\section{Formalism}

Let us consider the operators

\bea \label{1} 
A = {\hbar \over \sqrt{2m}}{d \over dx} + W(x)~,~ A^{\dag} = { -\hbar
\over \sqrt{2m}}{d \over dx} + W(x)~.
\eea 
From these two operators, we can construct two Hamiltonians $H_1, H_2$
given by 
\be\label{2}
H_1=A^{\dag}A\,,~~H_2=AA^{\dag}\,.
\ee
It is easily checked that
\be\label{3}
 H_1=-\frac{\hbar^2}{2m} \frac{d^2}{dx^2}+V_1(x)\,,~~~
V_1(x) = W^2(x) - {\hbar \over \sqrt{2m}}W^{\prime}(x)\,,
\ee
\be\label{4}
 H_2=-\frac{\hbar^2}{2m} \frac{d^2}{dx^2}+V_2(x)\,,~~~
V_2(x) = W^2(x) + {\hbar \over \sqrt{2m}}W^{\prime}(x)\,,
\ee
The quantity $W(x)$ is generally referred to as the superpotential 
in SUSY QM literature while $V_{1,2} (x)$ are termed as the
supersymmetric  partner potentials.  

 As we shall see, the energy eigenvalues, the wave functions and the S-matrices  of $H_1$
and $H_2$ are related. In particular, we shall show that if $E$ is the
eigenvalue of $H_1$ then it is also the eigenvalue of $H_2$ and vice a
versa. 
To that end notice that the energy eigenvalues of both $H_1$ and 
$H_2$ are positive semi-definite  ($E_n^{(1,2)} \geq 0$) . Further, 
the Schr\"{o}dinger
equation for $H_1$ is
\be\label{5} 
H_1 \psi_n^{(1)} = A^{\dag} A \psi_n^{(1)} =  E_n^{(1)}\psi_n^{(1)}\,. 
\ee 
On multiplying both sides of this equation by the operator $A$ from
the left, we get
\be\label{6} 
H_2 (A \psi_n^{(1)})  = A A^{\dag} A \psi_n^{(1)} =  E_n^{(1)}(A \psi_n^{(1)})~.
\ee 
Similarly, the Schr\"{o}dinger equation for $H_2$
\be\label{7} 
H_2 \psi_n^{(2)} = A A^{\dag}  \psi_n^{(2)} =  E_n^{(2)}\psi_n^{(2)}
\ee 
implies
\be\label{8} 
 H_1 (A^{\dag}  \psi_n^{(2)})  =  A^{\dag} A A^{\dag} \psi_n^{(2)} =  E_n^{(2)}
(A^{\dag}
\psi_n^{(2)})~. 
\ee 

Thus we have shown that if $E$ is an eigenvalue of the Hamiltonian
$H_1 (H_2)$ with eigenfunction $\psi$, then same $E$ is also the eigenvalue
of the Hamiltonian $H_2 (H_1)$ and the corresponding eigenfunction is
$A\psi (A^{\dag}\psi)$. 

The above proof breaks down in case $A\psi_0^{(1)} =0$, i.e. when the
ground state is annihilated by the operator $A$ (note that
the spectrum of both the Hamiltonians is nonnegative). Thus the
exact relationship between the eigenstates of the two Hamiltonians
will crucially depend on if $A\psi_0^{(1)}$ is zero or nonzero, i.e.
if the ground state energy $E_0^{(1)}$ is zero or nonzero.

{\bf $A\psi_0^{(1)} \ne 0$:} In this case the proof goes through for all the
states including the ground state and hence all the eigenstates of the
two Hamiltonians are paired, i.e. they are related by ($n=0,1,2,...$)
\be\label{9} 
E_n^{(2)} = E_{n}^{(1)} > 0\,, 
\ee
\be\label{10} 
\psi_n^{(2)} = [E_{n}^{(1)}]^{-1/2} A \psi_{n}^{(1)}~,
\ee
\be\label{11}
\psi_{n}^{(1)} 
= [E_{n}^{(2)}]^{-1/2} A^{\dag}  \psi_{n}^{(2)}~.
\ee 

{\bf $A\psi_0^{(1)} =0$:} 
In this case, $E_0^{(1)} =0 $ and this state is unpaired while all
other states of the two Hamiltonian are paired. It is then 
clear that the eigenvalues and eigenfunctions of the two
Hamiltonians $H_1$ and $ H_2$ are related by
$(n=0,1,2,...)$
\be\label{12}
 E_n^{(2)} = E_{n+1}^{(1)}, \hspace{.2in} E_0^{(1)} 
= 0~,
\ee
\be \label{13}
\psi_n^{(2)} = [E_{n+1}^{(1)}]^{-1/2} A \psi_{n+1}^{(1)}~,
\ee
\be \label{14}
\psi_{n+1}^{(1)} 
= [E_{n}^{(2)}]^{-1/2} A^{\dag}  \psi_{n}^{(2)}~.
\ee 

The equation $A\psi_0^{(1)}=0$ can be interpreted in the following two
different ways depending on whether the superpotential $W(x)$ or the
ground state wave function $\psi_0^{(1)}$ is known. In case $W(x)$ is
known then
one can solve the equation $A\psi_{0}^{(1)}=0$ by using eq.
(\ref{1}) and the ground state wavefunction of $H_1$ 
is given, in terms of the superpotential $W$ by
\be\label{15}
\psi_0^{(1)}= N \exp[-\frac{\sqrt{2m}}{\hbar} \int^{x} W(y) dy]\,.
\ee

Instead, if $\psi_0^{(1)}$ is known, then this equation gives us the 
superpotential $W$, i.e. 
\be\label{16}
W(x) = -\frac{\hbar}{\sqrt{2m}} \frac{\psi_0'^{(1)} (x)}{\psi_0^{(1)} (x)}
\ee

Several Comments are in order at this stage.

\begin{enumerate}

\item Notice that if $\psi_{n+1}^{(1)}$ ( $\psi_n^{(2)}$) of $H_1$ ($H_2$) is normalized
then the wave function  $\psi_n^{(2)}$ ($\psi_{n+1}^{(1)}$) 
is also normalized. Similar remarks are obviously also
valid for eigenfunctions in eqs. (\ref{10}) and (\ref{11}). 

\item The operator
$A$ ($A^{\dag}$) not only converts an eigenfunction of 
$H_1 (H_2)$ into an
eigenfunction of $H_2 (H_1)$ with the same energy, but it also destroys
(creates) an extra node in the eigenfunction.

\item Note that since the bound state wave functions
must vanish at $x=\pm \infty$  (or at the two ends) 
hence it is clear from eq. (\ref{15})
that $A\psi_0^{(1)}=0$ and $A^{\dag} \psi_0^{(2)}=0$ can never be
satisfied simultaneously and hence only one of the two ground state
energies can be zero. As a matter of convention we shall always 
choose $W$ in such a way that only $\psi_0^{(1)}$ (if at all) is
normalized so that $E_0^{(2)} >0$ (notice that $E_0^{(2)}=0$ requires
that $A^{\dag} \psi_0^{(2)}=0$, see eq. (\ref{7})).

\item In case $A\psi_0^{(1)}=0$, since the ground state wave function of $H_1$ is annihilated by the operator
$A$, this state has no SUSY partner.  Thus 
knowing all the eigenfunctions of $H_1$ we can determine the eigenfunctions
of $H_2$ using the operator $A$, and vice versa using $A^{\dag}$ we can
reconstruct all the eigenfunctions of $H_1$ from those of $H_2$ except
for the ground state. This is illustrated in Fig. \ref{fig:partners}.

%
%
\begin{figure}
 \epsfxsize=0.8\hsize
\centerline{\psfig{file=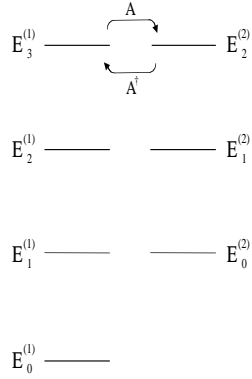,width=7.0truecm,height=7.0truecm}}
   \caption{ Energy levels of two (unbroken) supersymmetric partner potentials. 
The action of the operators $A$ and $A^\dag$ are displayed. The 
levels are degenerate except that $V_1$ has an extra state at
zero energy.}
   \label{fig:partners}
\end{figure}

\end{enumerate}

  The underlying reason for the degeneracy of the spectra of $H_1$ and $H_2$
can be understood most easily from the properties of the SUSY algebra. 
That is we can consider a
matrix SUSY Hamiltonian of the form
\be\label{17}
 H =  \left[\matrix{ H_1 &0\cr
0&H_2\cr}\right]
\ee
which contains both $H_1$ and $H_2$.  This matrix Hamiltonian is part of a
closed algebra which contains both bosonic and fermionic
operators with commutation and anti-commutation relations. We consider
the operators
\be \label{18}
Q= \left[\matrix{ 0 & 0\cr
A &0\cr}\right],
~~~Q^{\dag}= \left[\matrix{ 0 & A^{\dag}\cr
0&0\cr}\right]
\ee
in conjunction with $H$.  The following commutation and anticommutation 
relations
then describe the closed superalgebra $sl(1/1)$:

\bea\label{20}
[H,Q] &=&[H,Q^{\dag}] = 0 ~,\nonumber \\
\{Q,Q^{\dag}\} &=& H~~, \hspace{.2in}  \{Q,Q \} = \{Q^{\dag},Q^{\dag}\}=0~.
 \eea

The fact that the supercharges $Q$ and $Q^{\dag}$ commute with $H$ is
responsible for the  degeneracy in the spectra of $H_1$ and $H_2$.

Let us now try to understand as to when is SUSY spontaneously broken
and when does it remain unbroken. In this context, let us recall that
a symmetry of the Hamiltonian (or Lagrangian) can be spontaneously
broken if the lowest energy solution does not respect that symmetry, as for
example in a ferromagnet, where rotational invariance of the Hamiltonian
is broken by the ground state.  We can define the ground state in our system
by a two dimensional column vector:
\be\label{21}
\psi_0= \left[\matrix{ \psi_0^{(1)} \cr \psi_0^{(2)} \cr}\right]\,.
\ee
Then it is easily checked that in case $A\psi_0^{(1)} \ne 0$ then
$Q\psi_0 \ne 0$, $Q^{\dag}\psi_0 \ne 0$ 
and so SUSY is spontaneously broken 
while if $A\psi_0^{(1)} =0$ then $Q\psi_0=0, Q^{\dag}\psi_0 =0$ and 
SUSY remains unbroken. Unless stated
otherwise, throughout these
lectures we shall be discussing the case when SUSY remains unbroken.

Summarizing, we thus see that when SUSY is unbroken, then
starting from an exactly solvable potential $V(x)$ with $n$
bound states and ground state energy $E_0$, one has $V_1 (x) =
V(x)-E_0$ whose ground state energy is therefore 0 by construction.
Using the above formalism, one can then immediately obtain all the
$n-1$ eigenstates of $H_2$. One can now start from the exactly
solvable Hamiltonian $H_2$ and obtain all the $n-2$ eigenstates of
$H_3$. In this way, by starting from an exactly solvable problem with
$n$ bound states, one can construct new $n$ exactly solvable
potentials $H_2,H_3,...,H_{n+1}$ with $n-1, n-2,...,0$ bound states.

{\bf Illustration:} 
Let us look at a well known potential, namely the infinite square well
and determine its SUSY partner potential. Consider a particle of mass $m$ in an
infinite square well potential of width $L$ 
\bea\label{22}
V(x) &=& 0, \hspace{.5in} 0 \leq x \leq L ~, \nonumber \\
   &=& \infty, \hspace{.5in} -\infty < x <0~, x > L~.
\eea
The normalized ground state wave function is known to be
\be\label{23}
\psi_0^{(1)} = (2/L)^{1/2} \sin(\pi x/L), \hspace{.3in} 0 \leq x \leq L~,
\ee
and the ground state energy is 
\be
E_0= {\hbar^2 \pi^2 \over 2mL^2}  \, . 
\ee
Subtracting off the ground state 
energy so that the Hamiltonian can be factorized,
the energy eigenvalues of 
$H_1$ = $H- E_0$ are 
\be\label{24}
E_n^{(1)} = {n(n+2) \over 2mL^2} \hbar^2 \pi^2 \, , \ n = 0,1, 2, ... \, , 
\ee
and the normalized eigenfunctions of $H_1$ are (the same as those of
$H$), i.e. 
\be\label{25}
\psi_n^{(1)} = (2/L)^{1/2} \sin {(n+1)\pi x \over L}~,
\hspace{.5in} 0 \leq x \leq L ~.
\ee
The superpotential for this problem is readily obtained using eqs. (\ref{16})
and (\ref{23})
\be\label{26}
W(x) = - {\hbar \over \sqrt{2m}} {\pi \over L} {\rm cot} (\pi x /L)
\ee
and hence the supersymmetric partner potential $V_2$ is
\be\label{27}
V_2(x) = {\hbar^2 \pi^2 \over 2mL^2} [2~ {\rm cosec}^2 (\pi x /L) -1 ]~.
\ee
The wave functions for $H_2$ are obtained by applying the operator $A$
to the wave functions of $H_1$. In particular we find that the normalized
ground and first excited state wave functions are
\be\label{28}
\psi_0^{(2)} = -2\sqrt{\frac{2}{3L}} \sin^2 (\pi x /L) ,\hspace{.3in}
\psi_1^{(2)} = -\frac{2}{\sqrt{L}} \sin (\pi x /L)\sin (2\pi x /L)~.
\ee

\begin{figure}
\centerline{\psfig{file=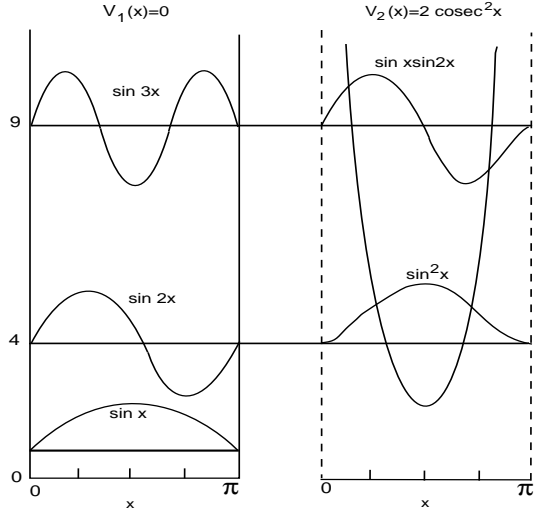,width= 7.0truecm, height =7.0truecm}}
\caption{The infinite square well potential $V=0$ of width $\pi$ and its
partner potential $V= 2 ~\rm{cosec}^2 x$ in units $\hbar=2m=1$}
\label{fig:partner}
\end{figure}

Thus we have shown using SUSY that two rather different
potentials corresponding to $H_1$ and $H_2$ have exactly the same spectra except
for the fact that $H_2$ has one fewer bound state.
 In Fig. \ref{fig:partner}
 we show
the supersymmetric partner potentials $V_1$ and $V_2$ and the first few
eigenfunctions. For  convenience we have chosen $L=\pi$ and $\hbar=2m=1$.

We can now start from $H_2$ and using its ground state wavefunction as
given by eq. (\ref{28}), the corresponding superpotential turns out to
be
\be\label{29}
W_2(x) = - {\hbar \over \sqrt{2m}} {2\pi \over L} {\rm cot} (\pi x
/L)\,,
\ee
and hence the partner potential $V_3 (x)$ turns out to be
\be\label{30}
V_3(x) = {\hbar^2 \pi^2 \over 2mL^2} [6~ {\rm cosec}^2 (\pi x /L) -4 ]~.
\ee
Using the above superpotential $W_2$, the ground state wavefunction of
$V_3 (x)$ is easily computed by using eq. (\ref{13})  
\be\label{31}
\psi_0^{(3)} (x)~ \propto~ \sin^3 (\pi x /L)\,.
\ee
In this way one can construct one (discrete) parameter family of
potentials given by ($p=0,1,2,...$) 
\be\label{32}
V_{p+1}(x) = {\hbar^2 \pi^2 \over 2mL^2} [p(p+1)~ {\rm cosec}^2 (\pi x
/L) -p^2 ],
\ee
and it is easily seen that its spectrum is given by
\be\label{33}
E_n^{(p+1)} = n(n+2p+2)\,,~~n=0,1,2,...\,,
\ee
while its eigenfunctions can be easily derived recursively from those
of infinite-square well. For example, its ground state wave function is
\be\label{34}
\psi_0^{(p+1)} ~\propto ~\sin^{(p+1)} (\pi x /L)\,.
\ee
In this way we have shown, how to generate a whole class of new 
solvable potentials
by starting from an analytically solvable problem. 

So far, we have explicitly included all factors like $\hbar,m$ etc.
However, from now onward, for simplicity (unless stated otherwise),
we shall work in units where $\hbar = 1,2m=1$.

{\bf Scattering:}  Supersymmetry also allows one to 
relate the reflection and transmission coefficients
in situations where the two partner potentials
have continuous spectra. In order for
scattering to take place in both of the partner potentials, it is necessary
that the potentials $V_{1,2}$ are finite as $x \rightarrow  - \infty $ or as 
$x \rightarrow  + \infty $ or both.  Let us define
\be\label{35}
W(x \rightarrow \pm \infty) \equiv W_{\pm}~.
\ee
Then it follows that
\be\label{36}
V_{1,2} \rightarrow W_{\pm}^2 \hspace{.5in} {\rm as} \hspace{.1in} x \rightarrow
\pm \infty \, .
\ee

Let us consider an incident plane wave $ e^{ikx}$ of energy $E$ coming from
the direction $ x \rightarrow - \infty$. As a result of scattering from the
potentials
$V_{1,2}(x)$ one would obtain transmitted waves $T_{1,2}(k) e^{ik'x}$
and reflected waves $R_{1,2}(k) e^{-ikx}$.
Thus we have
\bea\label{37}
\psi^{(1,2)}(k, x \rightarrow - \infty) && \rightarrow  e^{ikx} + R_{1,2}
e^{-ikx}~, \nonumber \\ \psi^{(1,2)}(k', x \rightarrow + \infty) && \rightarrow
T_{1,2} e^{ik'x}~,
\eea
where $ k$ and $k^{\prime}$ are given by
\be\label{38}
k= (E- W_{-}^2)^{1/2}~, \hspace{.2in} k^{\prime}= (E- W_{+}^2)^{1/2}~.
\ee
SUSY connects continuum wave functions  of $H_1$ and $H_2$ having the
same energy analogously to what happens in the discrete spectrum.
Thus using eqs. (\ref{13}) and (\ref{14}) we have the relationships:
\bea\label{39}
 e^{ikx} + R_{1} e^{-ikx} &=& N [ (-ik+W_{-}) e^{ikx} +(ik+W_{-}) e^{-ikx} R_2]
~, \nonumber \\
T_{1} e^{ik'x} &=&N[ (-ik'+W_{+}) e^{ik'x} T_2]~,
\eea
where $N$ is an overall normalization constant.  On equating terms with the same
exponent and eliminating $N$, we find:
\bea \label{40}
R_1(k) &=& \left( {W_{-} + ik \over W_{-} - ik }\right) R_2(k)~, \nonumber \\
T_1(k) &=& \left( {W_{+} - ik^{\prime} \over W_{-} - ik }\right) T_2(k)~.
\eea

A few remarks are in order at this stage.\\
(1) Clearly $ |R_1|^2 =|R_2|^2$ and  $ |T_1|^2 =|T_2|^2$, that is the
partner potentials  \index{SUSY!partner potentials} have identical reflection and transmission probabilities.\\
(2) $R_1(T_1)$ and $R_2(T_2)$ have the same poles in the complex plane
except that $R_1 (T_1)$ has an extra pole at $k=-iW_{-}$.  This pole is on
the positive imaginary axis only if $W_{-} <0$ in which case it corresponds to
a zero energy bound state.\\
(3) In the special case that $W_{+} =W_{-}$, we have that $T_1(k) = T_2(k)$.\\
(4) When $W_{-} =0$ then $R_1(k)= -R_2(k)$.\\
(5) For symmetric potentials, $W_{+} = - W_{-}$, and hence $k=k'$ so
that the relation between $T_1$ and $T_2$ is the same as that between
$R_1$ and $R_2$, i.e. $R_1(k)/R_2(k)=T_1(k)/T_2(k)$. 

{\bf Reflectionless Potentials:} 
It is clear from these remarks that if one of the partner potentials  is a
constant  potential (i.e. a free particle), then the other partner will be
of necessity reflectionless.  In this way we can understand the reflectionless
potentials of the form $ V(x) =  n(n+1)~ {\sech}^2 x $ which play a
critical role in understanding the soliton  solutions of the
Korteweg-de Vries (KdV)  equation. 
Let us consider the superpotential 
\be\label{41}
W(x) = B \ \rm{tanh}~  x~.
\ee
The two partner potentials are
\bea\label{42}
V_1 &=& B^2 - B \ (B+1) \rm{sech}^2  x~,
\nonumber \\   V_2 &=& B^2 - B(B-1) 
\rm{sech}^2~ x ~.
 \eea
We see that for $B=1$,  $ V_2(x)$ corresponds to
a constant potential (free particle) with no bound states, 
transmission coefficient $T_2 (k)=1$ and $R_2 (k) =0$. Hence the 
corresponding $V_1$ is a reflectionless
potential with precisely one bound state at $E_0^{(1)}=0$. Further by
using eq. (\ref{40}) it follows that its transmission coefficient 
is given by
\be\label{43}
T_1(k,B=1) =  \frac{(1 - ik)}{(-1 - ik)}~.
\ee

If instead we choose $B=2$ then one finds that $V_1 = 4-6 \rm{sech}^2
x$, while as seen above, $V_2 =4-2\rm{sech}^2 x$ is reflectionless potential
with one bound state!  Hence it follows that $V_1$ must be a
reflectionless potential with two bound states and 
using the SUSY machinery (eqs. (\ref{12}) to (\ref{15})), we can immediately
obtain the two eigenvalues of $V_1$ and the corresponding
eigenfunctions. They are
\be\label{43a}
E_0^{(1)} (B=2) = 0\,,~~ \psi_0^{(1)}~\propto \rm{sech}^2 x\,,
\ee
\be\label{43b}
E_1^{(1)} (B=2) = 3\,,~~ \psi_1^{(1)}~\propto \rm{sech} x\,\tanh x \,.
\ee
Further using eqs. (\ref{40}) and (\ref{43})
one can show that its transmission amplitude is given by
\be\label{44}
T_1(k,B=2) =  \frac{(2 - ik)(1-ik)}{(-2-ik)(-1 - ik)}~.
\ee
Generalization to arbitrary integer $B$ is now straight forward, i.e.
by choosing $B=3,4,...$, it is then easy to see that the one
(discrete) parameter family of reflectionless potentials 
is given by ($p=1,2,3,...$) 
\be\label{44a}
V_1 (x,B=p)=p(p+1) \rm{sech}^2 x-p^2 \,,
\ee
having $p$ bound states whose energy eigenvalues and
eigenfunctions can be recursively obtained by starting from free
particle and using the SUSY machinery. For example, its ground state
energy is zero while the ground and first excited state eigenfunctions
are
\be
\psi_0^{(1)} (x,B=p)~\propto~\sech^p x\,,~~\psi_1^{(1)}
(x,B=p)~\propto~\sech^{p-1} x\,\tanh x\,.
\ee
Further, its transmission coefficient
is given by
\be\label{45}
T_1 (k,p) =
\frac{(p-ik)(p-1-ik)...(1-ik)}{(-p-ik)(-p+1-ik)...(-1-ik)} 
= \frac{\Gamma(-p-ik) \Gamma(p+1-ik)}{\Gamma(-ik) \Gamma(1-ik)}\,.
\ee

{\bf SUSY in n-dimensions:} 
So far we have discussed SUSY QM on the full line $(-\infty \le x \le \infty)$.
Many of these results have analogs for the $n$-dimensional potentials with
spherical symmetry.  For example, for spherically
symmetric potentials in three dimensions one can make a partial wave
expansion in terms of the wave functions:
\be\label{46}
\psi_{nlm}(r,\theta,\phi) = {1 \over r} R_{nl}(r) Y_{lm} (\theta, \phi)~.
\ee
Then it is easily shown that the reduced radial wave function
$R_{nl}$ satisfies  
$(0 \le r \le \infty)$
\be\label{47}
-{d^2 R_{nl} (r) \over dr^2 } + \biggl[ V(r)+\frac{l(l+1)}{r^2}
 \biggr] R_{nl} (r) \equiv H R_{nl} (r) = E R_{nl} (r) \, .
\ee
Notice that this is a Schr\"odinger equation for 
an effective one dimensional potential which contains
the original potential plus an angular momentum barrier.  

As an illustration, let us now show how both the Coulomb and the oscillator
potentials can be cast in SUSY formalism. 

{\bf Oscillator Potential:} It is easily seen that in case we start
with 
\be\label{48}
W(r) = \frac{\omega r}{2} -\frac{l+1}{r}\,,
\ee
then the two supersymmetric partner potentials
are
\bea\label{49}
V_1 (r) &=& \frac{\omega^2 r^2}{4} +\frac{l(l+1)}{r^2}-(l+3/2)\omega
\nonumber \\
V_2 (r) &=& \frac{\omega^2 r^2}{4}
+\frac{(l+1)(l+2)}{r^2}-(l+1/2)\omega\,.
\eea
Thus we see that in the oscillator case, a given partial wave $l$ 
has as its SUSY partner the same oscillator potential but for the
$l+1$'th partial wave. In particular, the S and P-wave oscillator
potentials are the SUSY partners of each other. 

{\bf Coulomb Potential:} Let us start
with 
\be\label{50}
W(r) = \frac{e^2}{2(l+1)} -\frac{l+1}{r}\,,
\ee
then it is easily shown that the two supersymmetric partner potentials
are
\bea\label{51}
V_1 (r) &=& \frac{e^4}{4(l+1)^2} +\frac{l(l+1)}{r^2}-\frac{e^2}{r}
\nonumber \\
V_2 (r) &=& \frac{e^4}{4(l+1)^2} +\frac{(l+1)(l+2)}{r^2}-\frac{e^2}{r}
\nonumber \\
\eea
Thus we see that even for the Coulomb potential, a given partial wave $l$ 
has as its SUSY partner the same Coulomb potential but for the
$l+1$'th partial wave. 

As in the one dimensional case, we can also obtain the relationship
between the phase-shifts for the (radial)  partner potentials. The
asymptotic form of the radial wave function for the $l'$th partial
wave is
\be\label{52}
R (r,l) \rightarrow { 1 \over 2k^{\prime}}[S_{l}(k') e^{ik'r}- (-1)^{l}
e^{-ik'r}]~,
\ee
where $S_l$ is the scattering function for the $l'$th 
partial wave,
i.e. $S_l(k) = e^{i\delta_l(k)}$ and $\delta_l$ is the phase shift for
the l'th partial wave. Proceeding exactly as in the one dimensional
case (see eqs. (\ref{35}) to (\ref{40})), we obtain the following relationship
between the scattering functions (and hence phase shifts) for the two
partner potentials
\be \label{53}
S^{(1)}_l(k') = \left( {W_{+} - ik' \over W_{+} + ik' }\right) S^{(2)}_{l}(k')~.
\ee
Here $W_{+} = W(r \rightarrow \infty)$ and $k'=\sqrt{E-W_{+}^2}$. 

Before finishing this section, it is amusing to note that the famous
problem of charged particle in a uniform magnetic field can in fact be
cast in the language of SQM. Actually one can prove a stronger
result. In particular, one can show that not only
uniform but even for an arbitrary magnetic field, the Pauli equation
in two dimensions 
can be cast in the SQM formalism so long as the gyromagnetic
ratio $g=2$. 

Let us consider the special case when the motion of the charged
particle is in a plane perpendicular to the magnetic field. In this
case the Pauli Hamiltonian,for $g=2$ has the form
\be\label{54}
H_{Pauli} = \frac{1}{2m} \sum_{i=1}^{2} (p_i+\frac{e}{c}A_i)^2
+\frac{e\hbar}{2mc}B_3 \sigma_3\,.
\ee
One can show that in this case if we define
\bea
Q&&= {i \over \sqrt{2}} (Q^{(1)}+ i Q^{(2)})~, \nonumber \\
Q^{(1)} &&= {1 \over \sqrt{2}} \left[-(p_y+{e \over c} A_y) \sigma_1
+ (p_x +{e \over c}A_x) \sigma_2 \right]~, \nonumber \\
Q^{(2)} &&= {1 \over \sqrt{2}} \left[(p_x+{e \over c}A_x) \sigma_1
+ (p_y +{e \over c}A_y) \sigma_2 \right] ~,
\eea
then the Hermitian supercharges $Q^{(1)},
Q^{(2)}$
 and $H_{Pauli}$ satisfy the SUSY algebra
\be
\{Q^{(a)},Q^{(b)} \} = H_P \delta^{ab}~, ~~[H_P,Q^{(a)}]=0~, ~~a,b=1,2~.
\ee

\section{Shape Invariance and Solvable Potentials}

Using the ideas of SUSY QM developed in the last section and 
an integrability condition called the shape invariance 
condition, we now show that the operator method for the harmonic
oscillator can be generalized to the whole class of shape invariant potentials
(SIP) which include essentially all the popular,
analytically solvable potentials. Indeed, we shall see that for such
potentials, the generalized operator method quickly yields all the bound state
energy eigenvalues, eigenfunctions as well as the scattering matrix. It turns
out that this approach is essentially equivalent to Schr\"{o}dinger's method
of factorization \cite{sch31} 
although the language of SUSY is more appealing.

Let us now explain precisely what one means by shape invariance.
If the pair of SUSY partner potentials $V_{1,2}(x)$
as defined in eq. (\ref{3}) are similar in shape and differ only in
the parameters that appear in them, then they are said to be shape invariant.
More precisely, if the partner potentials $V_{1,2}(x;a_1)$ satisfy the
condition\cite{gan83}
\be \label{eq4.1}
V_2(x;a_1) = V_1(x;a_2) + R(a_1) \, ,
\ee
where $a_1$ is a set of parameters, $a_2$ is a function of $a_1$ (say
$a_2=f(a_1)$) and the remainder $R(a_1)$ is independent of $x$, then
$V_{1}(x;a_1)$ and $V_{2}(x;a_1)$ are said to be shape invariant. The shape
invariance condition (\ref{eq4.1}) is an integrability condition. Using
this condition and the hierarchy of Hamiltonians 
discussed in the previous section, one can easily obtain the energy
eigenvalues and the eigenfunctions of any SIP when SUSY is unbroken.

Let us start from the SUSY partner Hamiltonians 
$H_1$ and $H_2$ whose
eigenvalues and eigenfunctions are related by SUSY.
Further, since SUSY is unbroken we know that
\be \label{eq4.2}
E^{(1)}_0(a_1)=0,\quad \psi^{(1)}_0(x;a_1)
= N \exp\left[-\int^xW_1(y;a_1)dy\right] \, .
\ee
We will now show that the entire spectrum of $H_1$ can be very easily
obtained algebraically by using the shape invariance
condition (\ref{eq4.1}). As a first step, let us obtain the eigenvalue
and the eigenfunction of the first excited state of $H_1$. We start
from eq. (\ref{eq4.1}) and on adding kinetic energy operator to both the
sides, the shape invariance condition can also be written as  
\be \label{eq4.2a}
H_2(x;a_1) = H_1(x;a_2) + R(a_1) = A^{\dag}(x,a_2) A (x,a_2) +R (a_1)\, .
\ee
Since the two Hamiltonians differ by a constant, hence it is clear
that all their eigenvalues must differ from each other only by that
constant and further all their eigenfunctions must be proportional to
each other. In particular
\be\label{4.2b}
E_0^{(2)} (a_1) = E_0^{(1)} (a_2) +R(a_1)\,,~~\psi_0^{(2)} (x,a_1) 
~\propto~ \psi_0^{(1)} (x,a_2)\,.
\ee
But clearly $E_0^{(1)} (a_2)=0$ so long as the corresponding ground
state wave function 
\be
\psi_0^{(1)} (x,a_2) ~\propto~\exp[-\int^x
W(y,a_2)\,dy]\,, 
\ee
is an acceptable ground state wave function. And hence
using eqs. (\ref{4.2b}) and (\ref{12}) we obtain the eigenvalue of the
first excited state
\be\label{4.2c}
E_1^{(1)} (a_1) = E_0^{(2)} (a_1) = R(a_1)\,.
\ee
Similarly, using eqs. (\ref{4.2b}) and (\ref{13}) we can obtain the
eigenfunction of the first excited state
\be\label{4.2d}
\psi_1^{(1)} (x,a_1) ~\propto~A^{\dag} (x,a_1) \psi_0^{(2)} (x,a_1)
~\propto~A^{\dag} (x,a_1) \psi_0^{(1)} (x,a_2)\,.
\ee

This procedure is easily generalized and we can obtain the entire bound
state spectrum and the corresponding eigenfunctions of $H_1$. 
To that purpose, let us construct a series of
Hamiltonians $H_s$, $s=1,2,3$~. 
In particular, following the discussion of
the previous section, 
it is clear that if $H_1$ has $n$ bound states then one can
construct $n$ such Hamiltonians $H_2, H_3 \cdots H_{n+1}$ and the $p$'th
Hamiltonian $H_p$ will have the same spectrum as $H_1$ except that
the first $p-1$ levels of $H_1$ will be absent in $H_p$. On repeatedly
using the shape invariance condition (\ref{eq4.1}), it is then clear that
\be \label{eq4.3}
H_s = - {d^2\over dx^2} + V_1(x;a_s) + \sum^{s-1}_{k=1}R(a_k) \, ,
\ee
where $a_s=f^{s-1}(a_1)$ i.e. the function $f$ applied $s-1$ times. Let us
compare the spectrum of $H_s$ and $H_{s+1}$. In view of eqs. (\ref{eq4.1}) and
 (\ref{eq4.3}) we have
\bea \label{eq4.4}
H_{s+1} = -{d^2\over dx^2}+V_1 (x;a_{s+1})+\sum^s_{k=1} R(a_k) \nonumber\\
        =-{d^2\over dx^2}+V_2 (x;a_s)+\sum^{s-1}_{k=1} R(a_k) \, .
\eea
Thus $H_s$ and $H_{s+1}$ are SUSY partner Hamiltonians and hence have
identical bound state spectra except for the ground state of
$H_s$ whose energy is
\be \label{eq4.5}
E^{(s)}_0 = \sum^{s-1}_{k=1} R(a_k) \, .
\ee
This follows from eq. (\ref{eq4.3}) and the fact that $E^{(1)}_0=0$. On going
back from $H_s$ to $H_{s-1}$ etc, we would eventually reach $H_2$ and
$H_1$ whose ground state energy is zero and whose $n$'th level is
coincident with the ground state of the Hamiltonian
$H_n$. Hence the complete eigenvalue spectrum of $H_1$ is
given by
\be \label{eq4.6}
E^{(1)}_n(a_1) = \sum^n_{k=1} R(a_k); \hspace{.2in}  E^{(1)}_0(a_1)=0 \, .
\ee

As far as the corresponding eigenfunctions are concerned, we now show that, 
similar to the case of the one dimensional harmonic
oscillator, the bound state wave functions  $\psi^{(1)}_n(x;a_1)$ for
any shape invariant potential can also be easily obtained from its
ground state wave function $\psi^{(1)}_0(x;a_1)$ which in turn is
known in terms of the superpotential. This is possible
because the operators $A$ and $A^{\dag}$ link up the eigenfunctions of the
same energy for the SUSY partner Hamiltonians $H_{1,2}$. Let us
start from the Hamiltonian $H_s$ as given by eq. (\ref{eq4.3}) whose ground
state eigenfunction is then given by
 $\psi^{(1)}_0(x;a_s)$. On going from
$H_s$ to $H_{s-1}$ to $H_2$ to $H_1$ and using eq. (\ref{14}) we then find
that the $n$'th state unnormalized, energy eigenfunction
$\psi^{(1)}_n(x;a_1)$ for the original Hamiltonian $H_1(x;a_1)$ is
given by
\be \label{eq4.7}
\psi^{(1)}_n(x;a_1) ~\propto~ A^{\dag}(x;a_1)A^{\dag}(x;a_2)...A^{\dag}(x;a_n)
\,\psi^{(1)}_0(x;a_{n+1}) \, ,
\ee
which is clearly a generalization of the operator method of
constructing the energy eigenfunctions for the one dimensional
harmonic oscillator.

It is often convenient to have explicit expressions for the wave
functions. In that case, instead of using
the above equation, it is far simpler to use the identity
\be \label{eq4.8}
\psi^{(1)}_n(x;a_1) = A^{\dag}(x;a_1) \, \psi^{(1)}_{n-1}(x;a_2) \, .
\ee

Finally, in view of the shape invariance
condition (\ref{eq4.1}), the relation (\ref{40}) between scattering amplitudes
takes a particularly simple form
\be \label{eq4.9}
R_1(k;a_1)=\left({W_-(a_1)+ik \over W_-(a_1)-ik} \right) R_1(k;a_2) \, ,
\ee
\be \label{eq4.10}
T_1(k;a_1)=\left({W_+(a_1)-ik'\over W_-(a_1)-ik} \right) T_1(k;a_2) \, ,
\ee
thereby relating the reflection and transmission coefficients 
 of the same Hamiltonian $H_1$ at $a_1$ and $a_2(=f(a_1))$.

{\bf Illustrative Examples:} It may be worthwhile at this stage to
illustrate this discussion with concrete examples and obtain their
energy eigenvalues and eigenfunctions analytically. As a first
illustration, let us consider the partner potentials $V_{1,2}\, (x,B)$
as given by 
eq. (\ref{42}). It is easily seen that indeed these are shape
invariant. In particular we observe that
\be\label{60}
V_2 (x,B) = V_1 (x,B-1)+B^2-(B-1)^2\,,
\ee
so that in this particular case $a_1 = B, a_2 = B-1$ and $R(a_1)=B^2-
(B-1)^2 = a_1^2 -a_2^2$. Hence, from eq. (\ref{eq4.6}) it immediately
follows that the complete bound state spectrum of $V_1 (x,B)$ is given
by ($n=0,1,2,...$)
\be\label{61}
E_n^{(1)} (B) = \sum_{k=1}^{n} R(a_k) = a_1^2 -a_{n+1}^2 = B^2
-(B-n)^2\,.
\ee
It is really remarkable that in two lines one is able to get the entire
spectrum for this potential algebraically. What about the
eigenfunctions? Using the superpotential as given by eq. (\ref{41}) it
follows that
\be\label{62}
\psi_0^{(1)}(x,B) \propto sech^B (x)\,.
\ee
Clearly, this is an acceptable wave function so long as $B$ is any 
positive number. On the other hand, the first excited state
wavefunction is obtained by using the formula (\ref{eq4.7})
\be\label{63}
\psi_1^{(1)}(x,a_1=B) ~\propto~ A^{+} (x,B) \psi_0 (x,a_2=B-1)
~\propto~ [-\frac{d}{dx}+B\tanh x] \sech^{B-1} x ~\propto~
\sech^{B-1} x \tanh x \,.
\ee
Proceeding in this way, one can calculate all the bound state eigenfunctions
of $V_1 (x)$.

It is worth pointing out that here $B$ is any positive number and not
just an integer. Further it is also clear from here that the number of
bound states is $n+1$ in case $n <B \le n+1$. 
Note that only when $B$ is integer that 
$V_{1,2}$ are reflectionless potentials.

As a second example, it is easily checked  that the Coulomb and the oscillator
potentials (in fact in arbitrary number of dimensions) are also
examples of shape invariant potentials. For example, it is easy to see
that in the Coulomb case, the partner potentials as given by eq.
(\ref{51}) are shape invariant potentials (SIP) satisfying
\be\label{64}
V_2 (r,l,e)=V_1 (r,l+1,e)+[\frac{e^4}{4(l+1)^2}-\frac{e^4}{4(l+2)^2}]\,.
\ee
Thus in this case, $a_1 = l, a_2 = l+1$ and
$R(a_1)=\frac{e^4}{4a_1^2}-\frac{e^4}{4a_2^2}$ and hence the complete
spectrum of $V_1 (r,l,e)$ is given by ($n=0,1,2,...$)
\be\label{65}
E_n^{(1)} = [\frac{e^4}{4(l+1)^2} -\frac{e^4}{4(l+n+1)^2}]\,.
\ee

Similarly, the oscillator partner potentials (\ref{49}) are SIP
satisfying
\be\label{66}
V_2 (r,l,\omega) = V_1 (r,l+1,\omega)-(l+1/2) \omega+(l+5/2) \omega \,,
\ee
so that in this case $a_1=l, a_2 = l+1$ while $R(a_1) = 2\omega$ and
hence from eq. (\ref{eq4.6}) it immediately follows that the entire
spectrum of the oscillator potential $V_1 (r,l,\omega)$ is given by
\be\label{67a}
E_n^{(1)} = 2n \omega\,,~~n=0,1,2,...\,.
\ee

Let us now discuss the interesting but difficult question of the
classification of various solutions to the shape invariance condition
(\ref{eq4.1}). This is clearly an important problem because once such
a classification is available, then one can discover new
SIPs which are solvable by purely algebraic
methods. Unfortunately, the general problem is still unsolved, only 
two classes of
solutions have been found so far.
In the first class, the parameters $a_1$
and $a_2$ are related to each other by translation $(a_2=a_1+\alpha)$.

Remarkably enough, all well known analytically
solvable potentials found in most text books on nonrelativistic quantum
mechanics belong to this class. In the second class,
the parameters $a_1$
and $a_2$ are related to each other by scaling $(a_2=qa_1)$.

\subsection{Solutions Involving Translation}
\label{sec4.3.1}

We shall now point out the key steps that go into the classification
of SIPs 
in case  $a_2=a_1+\alpha$.
Firstly one notices the fact that the eigenvalue spectrum of the
Schr\"{o}dinger equation is always such that the $n$'th eigenvalue $E_n$
for large $n$ obeys the constraint
\be \label{eq4.31}
A/n^2 \leq E_n \leq B n^2 \, ,
\ee
where the upper bound is saturated by the infinite square well potential while
the lower bound is saturated by the Coulomb potential. Thus, for any
SIP, the structure of $E_n$ for large $n$ is
expected to be of the form
\be \label{eq4.32}
E_n \sim \sum_{\alpha}C_{\alpha} n^{\alpha}~,~~ -2\leq \alpha \leq 2 \, .
\ee
Now, since for any SIP, $E_n$ is given by eq. (\ref{eq4.6}), it follows
that if
\be \label{eq4.33}
R(a_k) \sim \sum_{\beta} k^{\beta}~,
\ee
then
\be \label{eq4.34}
-3 \leq \beta \leq 1~.
\ee

How does one implement this constraint on $R(a_k)$? While one has no
rigorous answer to this question, it is easily seen that a fairly
general factorizable form of $W(x;a_1)$ which produces the above
$k$-dependence in $R(a_k)$ is given by
\be \label{eq4.35}
W(x;a_1)=\sum^n_{i=1}[(k_i+c_i)g_i(x)+h_i(x)/(k_i+c_i)+f_i(x)]~,
\ee
where
\be \label{eq4.36}
a_1=(k_1,k_2...)~,~~ a_2 = (k_1+ \alpha, k_2+\beta...)~,
\ee
with $c_i, \alpha, \beta$ being constants. Note that this ansatz
excludes all potentials leading to $E_n$ which contain fractional
powers of $n$. On using the above ansatz for $W$ in the shape invariance
condition eq. (\ref{eq4.1}) 
one can obtain the conditions to be satisfied
by the functions $g_i(x), h_i(x), f_i(x)$. One important condition is
of course that only those superpotentials $W$ are admissible which give a square
integrable ground state wave function. 

In Table \ref{sip}, we give
expressions for the various shape invariant potentials $V_1(x)$,
superpotentials $W(x)$, parameters $a_1$ and $a_2$ and the corresponding
energy eigenvalues $E^{(1)}_n$. 
Except for first 3 entries of this
table, $W(x+x_0)$ is also a solution.

\begin{table}
\footnotesize
\caption{Shape invariant potentials with (n=1,2)  in which the parameters
$a_2$ and $a_1$ are related by translation 
$(a_2 = a_1 + \beta)$. The energy
eigenvalues and eigenfunctions
are given in units $\hbar = 2m = 1$. The constants $A, B, \alpha,
\omega, l$ are all taken $\geq
0$.  Unless otherwise stated, the range of potentials is $- \infty
\leq x \leq \infty, 0 \leq r
\leq \infty$. For spherically symmetric potentials, the full wave
function is $\varphi_{nlm}(r,
\theta, \phi) = \varphi_{nl}(r) Y_{lm}(\theta, \phi)$.}
\label{sip}
\bigskip
\begin{tabular}{llll} \hline
\\ Potential  & $W(x)$
  & 
$V_1(x;a_1)$  &
$a_1$  \\
\\
  \hline
\\ Shifted oscillator & $\frac{1}{2}\omega x-b$ &
$\frac{1}{4}\omega^2\left(x-\frac{2b}{\omega}\right)^2 - \omega/2$ &
$\omega$
\\ \\
\hline

3-D oscillator  & $\frac{1}{2}\omega r-\frac{(l+1)}{r}$

& $\frac{1}{4}\omega^2r^2+\frac{l(l+1)}{r^2} -(l+3/2)\omega$ { } { }

& $l$ \\ \\
  \hline

  Coulomb & $\frac{e^2}{2(l+1)}-\frac{(l+1)}{r}$ &
$-\frac{e^2}{r}+{l(l+1)\over r^2} +
\frac{e^4}{4(l + 1)^2}$ &
$l$
\\
\vspace*{2 mm} \\ \hline

Morse & $A - B$ exp $(-\alpha x)$ & $A^2+B^2 \exp(-2\alpha x) $

& A \\

& & { } $- 2B(A
+ \alpha/2) \exp(-\alpha x)   $\\
\hline

Scarf II & $A \tanh \alpha x$ + $B {\rm sech}~\alpha x$ & $A^2+(B^2-A^2-A
\alpha) {\rm sech}^2
\alpha x$  & A \\   { } { } (hyperbolic) & & { } { } $+B(2A+\alpha)
{\rm sech}~ \alpha x
\tanh \alpha x$  \\
  \hline

Rosen-Morse II & $A \tanh \alpha x + B/A$ & $A^2+B^2/A^2 - A(A +
\alpha) {\rm sech}^2 \alpha x$
& A \\  { } { } (hyperbolic) &$(B < A^2)$ & { } { } + 2$B$ tanh $\alpha x$ \\
\\
\hline

Eckart & $-A \coth \alpha r + B/A $ & $A^2+B^2/A^2 -2B$ coth $\alpha r$ & A \\
& { } { } $(B > A^2)$ & $+ A(A - \alpha)
{\rm cosech}^2 \alpha r$  &\\
\vspace*{2mm}
\\ \hline

Scarf I & $ A \tan \alpha x - B \ {\rm sec} \ \alpha x$ & $-A^2+(A^2+B^2-A
\alpha) {\rm sec}^2
\alpha x$ & A \\ 
(trigonometric) & { }  $\left(
-\frac{1}{2}\pi \leq \alpha x
\leq \frac{1}{2}\pi \right)$ &{ } $-B(2A - \alpha) \tan \alpha x\  {\rm sec} \
\alpha x$ \\
\\
\hline

P\"{o}schl-Teller & $A \coth \alpha r$ - $B$ {\rm cosech} $\alpha r$ &
$A^2+(B^2+A^2+A
\alpha) {\rm cosech}^2
\alpha r$ & A \\ { }  & { } { } $( A < B)$ & $- B(2A+
\alpha)$ coth $\alpha r$ cosech
$\alpha r$ \\
\\ \hline

Rosen-Morse I & $-A$ cot $\alpha x - B/A$ & $A(A - \alpha)$cosec$^2
\alpha x + 2B$
cot $ \alpha x$ & A \\ (trigonometric)& { } $(0 \leq \alpha x
\leq \pi)$ & { } $ -A^2 +
B^2/A^2$
\vspace*{1 mm}\\
\\
\hline
\end{tabular}
\end{table}

\begin{table}
\footnotesize
 Note that the
wave functions for the
first four potentials (Hermite and Laguerre polynomials) are special
cases of the confluent hypergeometric function while the rest (Jacobi
polynomials) are special
cases of the hypergeometric function.  
In the table
$s_1=s-n+a$, $s_2=s-n-a$,
$s_3=a-n-s$, $s_4=-(s+n+a)$.
\vspace{.25in}
\\
\begin{tabular}{llll} \hline
\\
$a_2$  & Eigenvalue $E_n^{(1)}$
  & Variable $y$ & Wave function $\psi_n(y)$  \\
\\
  \hline
\\
$\omega$ & $n\omega$ & $y = \sqrt{{\omega \over 2}}
\left( x-\frac{2b}{\omega}
\right) $ & exp
$\left( -\frac{1}{2}y^2\right) H_n(y)$\\
\\
\hline

$l + 1$ & $2n\omega$ & $y = \frac{1}{2}\omega r^2$ & $y^{(l+1)/2}$
exp$\left( - \frac{1}{2}y
\right) L_n^{l+1/2} (y)$ \\
\\
\hline

$l + 1$ & $\frac{e^4}{4(l + 1)^2} - \frac {e^4}{4(n + l
+ 1)^2}$ & $y =
\frac{re^2}{(n + l + 1)}$ &
$y^{(l+1)}$ exp$\left( -
\frac{1}{2}y
\right) L_n^{2l+1} (y)$ \\
\\
\hline

$A - \alpha$ & $A^2 - (A - n\alpha)^2$ & $y = {2B \over \alpha} e^{-\alpha
x}$, &   $y^{s-n}$ exp$\left(-\frac{1}{2} y\right) L_n^{2s-2n}(y)$ \\
& { } & $s = A/\alpha$\\
\hline

$A - \alpha$ & $A^2 - (A - n\alpha)^2$ & $y$ = sinh $\alpha x$, &
  $i^n(1 + y^2)^{-s/2} e^{-\lambda \tan^{-1} y }$ \\
& { } &  $s = A/\alpha, \lambda = B/\alpha$ &  $\times
P_n^{(i\lambda-s-1/2,-i\lambda-s-1/2)} (iy)$\\
\hline

$A - \alpha$ & $A^2 - (A - n\alpha)^2 $ & $y$ = tanh $\alpha x$, &
  $(1 - y)^{s_1/2} (1 + y)^{s_2/2}$  \\
&  $- B^2/(A - n\alpha)^2$ & $ s = A/\alpha, \lambda =
B/\alpha^2$ & 
$\times P_n^{(s_1, s_2)} (y)$\\
&  $+ B^2/A^2$ & $ a = \lambda/(s-n)$\\
\hline

$A + \alpha$ & $A^2 - (A + n\alpha)^2 $ & $y$ = coth $\alpha r$, &
  $(y - 1)^{s_3/2} (y + 1)^{s_4/2}$  \\
& $-B^2 /(A + n\alpha)^2$ &  $s = A/\alpha, \lambda =
B/\alpha^2$ & 
$\times P_n^{(s_3,s_4)} (y)$\\
& $+ B^2/A^2$&  $ a = \lambda/(n+s)$\\
\hline

$A + \alpha$ & $(A + n\alpha)^2 - A^2$ & $y$ = sin $\alpha r$, &
  $(1 - y)^{(s - \lambda)/2} (1 + y)^{(s + \lambda)/2}$  \\
& { }  & $s = A/\alpha, \lambda = B/\alpha$ & $\times P_n^{(s - \lambda -
1/2, s + \lambda - 1/2)} (y)$
\\ \\
  \hline

$A - \alpha$ & $A^2 - (A - n\alpha)^2 $ & $y$ = cosh $\alpha r$, &
  $(y - 1)^{(\lambda - s)/2} (y + 1)^{-(\lambda + s)/2}$  \\
& { }  &  $s = A/\alpha, \lambda = B/\alpha$ & $\times P_n^{(\lambda -s - 1/2,
- \lambda - s - 1/2)} (y)$
\vspace*{3 mm}\\
\hline

$A + \alpha$ & $(A + n\alpha)^2 - A^2 $ & $y$ = $i$cot $\alpha x$, &
  $(y^2 - 1)^{-(s+n)/2}$ exp $(a \alpha x)$  \\
&  $- B^2/(A + n\alpha)^2$ & $ s = A/\alpha, \lambda =
B/\alpha^2$ & $\times
P_n^{(-s - n +ia, - s - n - ia)} (y)$ \\
&$+ B^2/A^2 $ &  $a = \lambda/(s + n)$ \\
\hline

\end{tabular}
\end{table}

Several remarks are in order at this time.

\begin{enumerate}
\item Throughout this section we have used the convention of
$\hbar = 2m=1$. It would naively appear that if we had not put $\hbar
=1$, 
then the shape invariant potentials as given in Table \ref{sip} would all
be $\hbar$-dependent. However, it is worth noting that in each and
every case, the $\hbar$-dependence is only in the constant multiplying the
$x$-dependent function so that in each case we can always redefine the
constant multiplying the function and obtain an $\hbar$-independent
potential. For example, corresponding to the superpotential given by
eq. (\ref{41}), let us discuss a more general potential given by
\be\label{67}
W(x) = B \tanh (\alpha x+x_0)\,,
\ee
so that the corresponding
$\hbar$ and $m$-dependent partner potentials are given by
\bea  \label{eq4.53}
&&V_1(x,B)= W^2 -\frac{\hbar}{\sqrt{2m}} \frac{dW}{dx} =
B^2-B(B+\frac{\hbar \alpha}{\sqrt{2m}}) \sech^2 (\alpha x+x_0)
\nonumber  \\
&&V_2(x,A)= W^2 +\frac{\hbar}{\sqrt{2m}} \frac{dW}{dx} =
B^2-B(B-\frac{\hbar \alpha}{\sqrt{2m}}) \sech^2 (\alpha x+x_0)\,,
\eea
which are shape invariant.
On redefining
\be \label{eq4.54}
B(B+\frac{\hbar\alpha}{\sqrt{2m}}) = a ~;
\ee
where $a$ is $\hbar$ and $m$-independent parameter, we then have
a $\hbar$-independent potential $V(x) = V_1 (x) -B^2 = -a \sech^2
(\alpha x+x_0)$.
\item It may be noted that the Coulomb
as well as the harmonic oscillator
potentials in $n$-dimensions are also shape invariant potentials\index{potentials!shape invariant}.
\item What we have shown here is that shape invariance is a sufficient
condition for exact solvability. But is it also a necessary condition?
The answer is clearly no. Firstly,  
it has been shown that the solvable Natanzon
potentials
 are in general not shape invariant. However, for
the Natanzon potentials, the energy eigenvalues and wave functions are known
only implicitly.  Secondly there are various methods one of 
which we will discuss
below of finding potentials which are  strictly isospectral to the SIPs. These
are not SIPs but  for all of  these potentials, unlike the Natanzon case, the
energy eigenvalues and eigenfunctions are known in a closed form.
\end{enumerate}

{\bf Solutions Involving Scaling}

From 1987 until 1993 it was believed that the only shape invariant
potentials were those given in Table \ref{sip} and that there were no more shape
invariant potentials. However, starting in 1993, a
huge class of new shape invariant potentials have been discovered. It
turns out that for many of these new shape
invariant potentials, the parameters $a_2$ and $a_1$ are related by
scaling $(a_2=qa_1, 0 < q < 1)$ rather than by translation, a
choice motivated by the recent interest 
in $q$-deformed Lie algebras. 
Many of these potentials are reflectionless and have an
infinite number of bound states. So far, none of these potentials have
been obtained in a closed form but are obtained only in a series form.

\section{Strictly Isospectral Hamiltonians and SUSY}

Given any potential $V(x)$, the corresponding
bound state energy eigenvalues and the reflection and transmission
coefficients $T(k), R(k)$ 
are unique. But is the converse also true? In particular,
given the entire bound state spectrum $E_n$ and $T(k),R(k)$ at all
energies, is the potential uniquely defined? It turns out that the
answer to this question is {\it no}. This is of course well known for
a long time from the inverse scattering approach. In particular, it
turns out that if a potential holds $n$ bound states then there exist
$n$ continuous parameter family of strictly isospectral potentials
(i.e. potentials having same energy eigenvalues and same reflection and
transmission coefficients at all energies). 
These isospectral families are closely connected to multi-soliton
solutions of nonlinear integrable systems. 

In this section we approach this problem within the SQM formalism and
show that given any potential with at least one bound state, one can
very easily construct one continuous parameter family of strictly 
isospectral potentials having
same $E_n,R(k),T(k), so long as its ground state wave function is
known$. Of course this was known for a long time from
inverse scattering but 
the Gelfand-Levitan approach to finding them is
technically much more complicated than the supersymmetry approach described
here.

In SQM, the inverse scattering question can be posed as follows. Once
a superpotential $W(x)$ is given then the corresponding partner
potentials $V_{1,2} (x)$ are unique. But is the converse also true? In
particular, for a given $V_2(x)$ is the corresponding $W(x)$ and hence 
$V_1 (x)$ unique? 
In other words,  what are the various 
possible superpotentials  $\hat{W}(x)$ other than $W(x)$ satisfying
\be \label{5.1}
V_2(x) = \hat{W}^2(x) + \hat{W}^\prime(x)~.
\ee
If there are new solutions, then one would obtain new potentials
$\hat{V}_1(x) = \hat{W}^2 - \hat{W}^\prime$ which would be isospectral to
$V_1(x)$.    To find the
most general solution, let 
\be
\hat{W}(x) = W(x) + \phi(x)~,
\ee
in eq. (\ref{5.1}).  We then find that $y(x) = \phi^{-1}(x)$ satisfies the
Bernoulli equation
\be
y^\prime(x) = 1 + 2Wy~,
\ee
whose solution is
\be
\frac{1}{y(x)} = \phi(x) = \frac{d}{dx}~\ln [{\cal I}(x) + \lambda]~.
\ee
Here
\be \label{5.5}
{\cal I}(x) \equiv \int_{-\infty}^{x} [\psi^{(1)}_0 (x^\prime)]^2\, dx^\prime~,
\ee
$\lambda$ is a constant of integration and $\psi^{(1)}_0(x)$ is the
normalized ground state wave function of $V_1(x) = W^2(x) - W^\prime(x)$.
Thus the most general $\hat{W}(x)$ satisfying eq. (\ref{5.1}) is given by
\be\label{5.5a}
\hat{W}(x,\lambda) = W(x) + \frac{d}{dx} \ln [{\cal I}(x) + \lambda]~,
\ee
so that the one parameter family
 of potentials
\be\label{5.6}
\hat{V}_1(x,\lambda) = \hat{W}^2(x) - \hat{W}^\prime(x) = V_1(x) - 2 \frac{d^2}
{dx^2} \ln\, [{\cal I}(x) + \lambda]~,
\ee
have the same SUSY partner $V_2(x)$. 

The corresponding normalized ground state wave functions 
are
\be \label{5.7}
\hat{\psi}^{(1)}_0(x,\lambda)= \frac{\sqrt{\lambda(1+\lambda)}
\,\psi^{(1)}_0
(x)}{{\cal I}(x)+\lambda}\,,
\ee
while the excited state wave functions are easily obtained by using
relation (\ref{14}), i.e.
\be\label{5.7a}
\hat{\psi}^{(1)}_{n+1} (x,\lambda) = [-\frac{d}{dx}+W(x,\lambda)]
\,\psi^{(2)}_n (x) \,.
\ee

Several comments are in order at this stage.

\begin{enumerate}

\item Note that this family contains the original potential $V_1$.  This
corresponds to the choices $\lambda \rightarrow \pm \infty$.

\item The fact that the potentials $V_1 (x,\lambda)$ are strictly
isospectral, i.e. have same $E_n,R(k),T(k)$ can be seen as follows.
Firstly, since all of them have the same partner potential $V_2 (x)$,
hence it follows that they must have the same spectrum. Secondly,
the reflection and transmission coefficients  for these potentials
must be related to those of $V_2 (x)$ by the formula (\ref{40}).
But from eqs. (\ref{5.5}) and (\ref{5.5a}) it is easy to show that
$\hat{W(x=\pm \infty,\lambda)}  =W_{\pm}$
and hence the transmission and reflection coefficients for the entire
one parameter family of potentials are identical.

\item Note that while $E_n,R(k),T(k)$ are the same for the entire
family of potentials, the wave functions for all these potentials are
different. As is well known from inverse scattering, a potential is
uniquely fixed only when one specifies $E_n,R(k),T(k)$ and
normalization constant of one of the eigenfunction.

\item  Since $\cal{I}$ varies from $0$ to $1$ hence if we choose 
the continuous parameter $\lambda$ so that either $\lambda > 0$ or 
$\lambda < -1$ then we ensure that the corresponding ground state
wavefunctions $\psi_1 (x,\lambda)$ are nonsingular.   

\item The one parameter family of
strictly isospectral potentials $\hat{V}_1 (x,\lambda)$ can be shown to satisfy infinite
number of conserved charges $Q_i$ (i=1,2,3,...). Two of these are 
\bea\label{5.8}
Q_1 &=& \int_{-\infty}^{\infty} V(x,\lambda)~dx \nonumber\\
Q_2 &=& \int_{-\infty}^{\infty} x\,V(x,\lambda)~dx\,,
\eea
which satisfy $\frac{dQ_{1,2}}{d\lambda}=0$ as can be easily
checked by using eqs. (\ref{5.5a}) to (\ref{5.7}). The conserved
charge $Q_2$ implies that the one continuous parameter family of 
potentials are such that if one plots these potentials as a function
of $x$, then the area under the curve is the same for the entire
family of potentials. Similarly, all the other conservation laws 
tighten these potentials.

\end{enumerate}

\begin{figure}
\epsfxsize=0.8\hsize
\centerline{\psfig{file=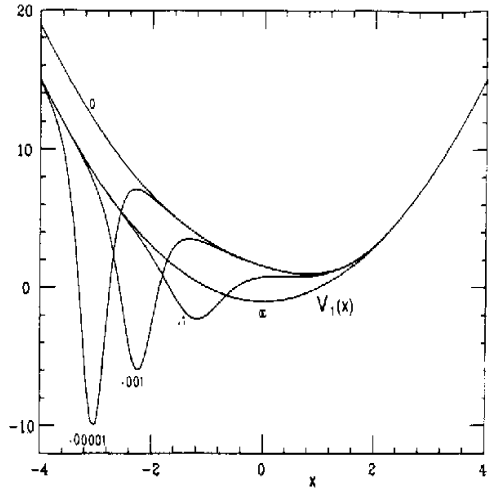,width=7.0truecm,height=7.0truecm}}
\caption{Selected members of the family of potentials with energy
spectra identical to the one dimensional harmonic oscillator with $\omega=2$.
The choice of units is $\hbar=2m=1$. The curves are labeled by the value
of $\lambda$, and cover the range $0<\lambda\le\infty$. The curve
$\lambda=\infty$ is the one dimensional harmonic oscillator.
The curve marked
$\lambda=0$ is known as the Pursey potential  and has one bound
state less than the oscillator.}
\label{fig:Pursey}
\end{figure}

\begin{figure}
\epsfxsize=0.8\hsize
\centerline{\psfig{file=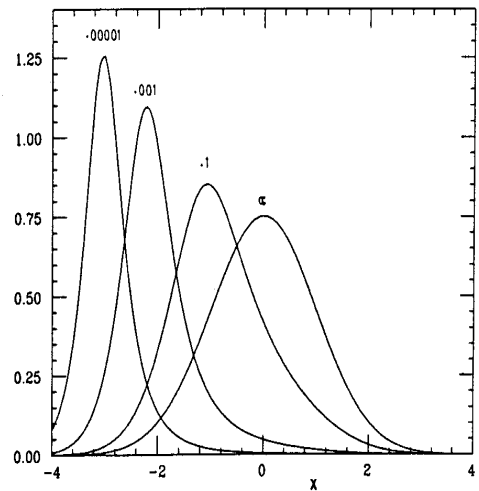,width=7.0truecm,height=7.0truecm}}
\caption{Ground state wave functions for all the potentials shown
in Fig. \ref{fig:Pursey}, except the Pursey potential.}
\label{fig:noPursey}
\end{figure}

To elucidate this discussion, it may be worthwhile to explicitly construct
the one-parameter family of strictly isospectral potentials corresponding
to the one dimensional harmonic oscillator. In this case
\be
W(x) = \frac{\omega}{2}x~,
\ee
so that
\be
V_1(x) = \frac{\omega^2}{4} x^2 - \frac{\omega}{2}~.
\ee
The normalized ground state eigenfunction of $V_1(x)$ is
\be
\psi^{(1)}_0(x) = \left(\frac{\omega}{2\pi}\right)^{1/4}\exp(-\omega{x^{2}}/4)~.
\ee
Using eq. (\ref{5.5}) it is now straightforward to compute the corresponding
${\cal I}(x)$.  We get
\be
{\cal I}(x)=1-\frac{1}{2}~{\rm erfc} \left(\frac{\sqrt{\omega}}{2} x\right)~;~
{\rm erfc}(x) = \frac{2}{\sqrt{\pi}}~\int_x^\infty~e^{-t^2} dt~.
\ee
Using eqs. (\ref{5.6}) and (\ref{5.7}),
one obtains the one parameter family of
isospectral potentials and the corresponding  
wave functions.  In Figs.
\ref{fig:Pursey} and \ref{fig:noPursey}  , we have plotted some of the
potentials and the corresponding ground state wave
 functions for the case $\omega = 2$.

We see that as $\lambda$ decreases from $\infty$ to 0,
$\hat{V}_1$ starts developing a minimum which shifts towards $x = -\infty$.
Note that as $\lambda_1$ finally becomes zero this attractive
potential well is lost
and we lose a bound state. The remaining potential is called the Pursey
potential  $V_P(x)$. The general formula for
$V_P(x)$ is obtained by
putting $\lambda=0$ in eq. (\ref{5.6}). An analogous situation occurs in
the limit $\lambda=-1$, the remaining potential being the Abraham-Moses
potential. 

\section{Supersymmetry Inspired WKB Approximation}

We will  
describe here a recent extension of the semiclassical approach inspired by
supersymmetry called 
the supersymmetric WKB (SWKB) method. It turns out that for
many problems the SWKB method gives better accuracy than the WKB method.  
Further, 
we discuss and prove the remarkable result that the 
lowest order SWKB
approximation gives exact energy 
eigenvalues for all SIPs with translation which include essentially
all analytically solvable potentials discussed in most text books on
QM. In this section we do not assume $\hbar=2m=1$ but explicitly keep
all the factors of $\hbar$ and $m$.

The semiclassical WKB
approximation for one dimensional potentials with two classical
turning points is discussed in most quantum mechanics textbooks.
The lowest order WKB quantization condition is given by
($n=0,1,2,...$)
\be\label{7.13}
\int^{x_R}_{x_L} dx \sqrt{2m[E-V(x)]} = (n+ 1/2) \hbar \pi \, .
\ee
In the special case of the one dimensional harmonic
oscillator and the Morse potential,
it turns out that the lowest order WKB approximation
(\ref{7.13}) is in fact exact
and further, the higher order corrections are all zero.

About twenty years ago, combining the ideas
of SUSY with the lowest order WKB method, 
the lowest order SWKB quantization condition was obtained 
in case SUSY is unbroken and it was shown that it yields energy
eigenvalues which are not only
accurate for large quantum numbers $n$  but which, 
by construction, are also exact
for the ground state $(n=0)$. We shall now discuss the SWKB 
quantization in detail.

 For the potential $V_1(x)$ corresponding to the
superpotential $W(x)$, the lowest order WKB quantization condition
(\ref{7.13}) takes the form
\be \label{3.2}
\int_{x_L}^{x_R} \sqrt{2m \bigg [E_n^{(1)}
-W^2(x)+\frac{\hbar}{\sqrt{2m}}W'(x) \bigg ]}~dx
        =(n+1/2)\hbar\pi.
\ee
Let us assume that the superpotential $W(x)$ is formally $O(\hbar^0)$.
Then, the $W'$ term is clearly
$O(\hbar)$. Therefore, expanding the left hand side in powers of $\hbar$ gives
\be  \label{3.3}
\int_{a}^{b} \sqrt{2m[E_n^{(1)}-W^2(x)]}~dx+\frac {\hbar}{2}
\int_{a}^{b} \frac{W'(x) ~dx}{\sqrt{E_n^{(1)}-W^2(x)}}+...
=(n+1/2)\hbar\pi,
\ee
where $a$ and $b$ are the turning points \index{turning points} defined by $E_n^{(1)}=W^2(a)
=W^2(b)$. The $O(\hbar)$ term in eq. (\ref{3.3}) can be integrated
easily to yield
\be \label{term}
\frac{\hbar}{2} \sin^{-1}\left[\frac{W(x)}{\sqrt{E_n^{(1)}}}\right]_{a}^{b}.
\ee
In the case of unbroken SUSY 
the superpotential $W(x)$ has
opposite signs at the two turning points, that is
\be \label{oppsign}
-W(a)=W(b)=\sqrt{E_n^{(1)}}~.
\ee
For this case, the $O(\hbar)$ term in (\ref{term}) exactly gives
$\hbar\pi/2$, so that to leading order in $\hbar$ the SWKB quantization
condition 
when SUSY is unbroken is
\be\label{3.4}
\int_{a}^{b} \sqrt{2m[E_n^{(1)}-W^2(x)]}~dx
=n\hbar\pi,~~n=0,1,2,...~.
\ee
Proceeding in the same way, the SWKB quantization condition
for the potential $V_2(x)$ turns out to be
\be\label{3.5}
\int_{a}^{b} \sqrt{2m[E_n^{(2)}-W^2(x)]}~dx
=(n+1)\hbar\pi,~~n=0,1,2,...~.
\ee

Some remarks are in order at this stage.

(i) For $n=0$, the turning points $a$ and $b$ in eq. (\ref{3.4}) are
coincident and $E_0^{(1)} = 0$. Hence the SWKB condition is exact by
construction for the ground state energy of the
potential $V_1(x)$.

(ii) On comparing eqs. (\ref{3.4}) and (\ref{3.5}), it follows that the
lowest order SWKB quantization condition preserves the SUSY
level degeneracy i.e. the approximate energy eigenvalues computed from the
SWKB quantization conditions for $V_1(x)$ and $V_2(x)$ satisfy the exact
degeneracy relation $E_{n+1}^{(1)}=E_n^{(2)}$.

(iii) Since the lowest order SWKB approximation is not only
exact, as expected, for large $n$,
but is also exact by construction for $n=0$, hence, unlike the ordinary
WKB approach, the SWKB eigenvalues are constrained to be accurate at both
ends, at least when the spectrum is purely discrete.
One can thus reasonably expect better results than the WKB scheme even when
$n$ is neither small nor very large.

(iv) For spherically symmetric potentials, unlike the conventional WKB
approach, in the SWKB case one obtains the correct threshold behaviour
without making any Langer-like correction. This happens because, in this approach
\be
S_0 \sim  (E-W^2)^{1/2} \ \stackrel{r \rightarrow 0}{\sim}
\ -i\hbar (l+1)/r \, ,
\ee
so that
\be
\psi(r) \sim
\exp \biggl[\frac{i}{\hbar} \int^{r} S_0 dr \biggr]
\ \stackrel{r \rightarrow 0}{\sim} \ r^{l+1} \, .
\ee

\subsection{Exactness of the SWKB Condition for Shape Invariant Potentials}

In order to determine the accuracy of the  SWKB quantization condition as given
by eq. (\ref{3.4}), researchers first 
obtained  the SWKB bound state spectra of several analytically
solvable potentials. Remarkably they found that  the lowest order SWKB
condition gives the exact eigenvalues for all 
SIPs with translation!
Let us now prove this result.

  Recall that the shape invariance condition eq. (\ref{eq4.1})  on
the partner potentials is
\[  
V_2 (x,a_1) = V_1 (x,a_2) +R(a_1) \, ,
\]
where $a_1$ is a set of parameters, $a_2$ is a function of $a_1$ (say
$a_2 = f(a_1)$) and the remainder $R(a_1)$ is independent of $x$.

In Section 3, we showed using factorization and the Hamiltonian hierarchy 
that the general expression for the $s$'th  Hamiltonian was given by 
\[
 H_{s} = -\frac{\hbar^2}{2m} \frac{d^2}{dx^2} +V_1 (x,a_s)
+\sum_{k=1}^{s-1} R(a_k)~,
\]
where $a_s = f^{s-1} (a_1)$ i.e. the function $f$ applied $s-1$ times.

The proof of the exactness of the bound state spectrum eq. (\ref{eq4.6}) in the
lowest order SWKB approximation now follows from the fact that
the SWKB condition (\ref{3.4}) preserves (a) the level
degeneracy and (b) a vanishing ground state energy eigenvalue.
For the hierarchy of Hamiltonians $H^{(s)}$ as given by eq. (\ref{eq4.3}),
 the SWKB
quantization condition takes the form
\be \label{3.10}
\int\sqrt{2m \bigg [E_n^{(s)}-\sum_{k=1}^{s-1}R(a_k)-W^2(a_s;x) \bigg ]}~dx
=n\hbar\pi~.
\ee
Now, since the SWKB quantization condition is exact
for the ground state energy when SUSY is unbroken, hence
\be
E_0^{(s)}=\sum_{k=1}^{s-1}R(a_k)
\ee
must be exact for Hamiltonian $H^{(s)}$
as given by eq. (\ref {3.10}).
One can now go back in sequential manner from $H^{(s)}$ to $H^{(s-1)}$ to
$H^{(2)}$ and $H^{(1)}$ and use the fact that the SWKB
method preserves the level degeneracy $E_{n+1}^{(1)}=E_n^{(2)}$.
On using this relation $n$ times, we find that for all SIPs,
the lowest order SWKB condition
gives the exact energy eigenvalues.

This is a substantial improvement over the usual WKB formula
eq. (\ref{7.13}) which is not exact for most SIPs. Of
course, one can artificially restore exactness by ad hoc Langer-like
corrections. However, such modifications are unmotivated
and have different forms for different potentials.
Besides, even with such corrections,
the higher order WKB contributions are non-zero for most of these
potentials.

What about the higher order SWKB contributions? Since the lowest order SWKB
energies are exact for shape invariant potentials,
it would be nice to check that higher
order corrections vanish order by order in $\hbar$. By starting from the
higher order WKB formalism, one can readily develop the higher order
SWKB formalism. It has been explicitly checked for all
known SIPs (with translation) that up to $O(\hbar^6)$
there are indeed no corrections.. This
result can be extended to all orders in $\hbar$.

Let us now compare the merits of the WKB and SWKB methods.  For
potentials for which the ground state wave function (and hence the
superpotential $W$) is not known, clearly the WKB approach is preferable, since
one cannot directly make use of the SWKB quantization condition (\ref{3.4}).
On the other hand, we have already seen that for shape invariant potentials,
SWKB is clearly superior. An obvious interesting question is to compare
WKB and SWKB for
potentials which are not shape invariant but for whom the ground state
wave function is known. An extensive study of several potentials 
indicate that by and
large, SWKB does better than WKB in case the ground state wave function
and hence the superpotential $W$ is known. These studies also support the
conjecture that shape invariance is perhaps a necessary condition so that
the lowest order SWKB reproduce the exact bound state spectrum.

\section{New Periodic Potentials From Supersymmetry}

So far we have considered potentials which have discrete or discrete
plus continuous
spectra and by using SUSY QM methods we have
generated new solvable potentials. In this
section we extend this
discussion to periodic potentials and their
band spectra. The
importance of this problem can hardly be overemphasized. For example,
the energy spectrum of electrons on a lattice is of central importance in
condensed matter physics. In particular, knowledge of the existence and
locations of band edges and band gaps
determines many physical properties of these systems.
Unfortunately, even in one dimension, there are very few analytically solvable
periodic potential problems. We show in this section that  SQM allows us
to enlarge this class of solvable periodic potential problems. 

We start from the 
Hamiltonians $H_{1,2}$ in which 
the SUSY partner
potentials $V_{1,2}$ are periodic nonsingular potentials
with a period $L$. 
 In view of the periodicity, one  seeks
solutions of the Schr\"odinger equation subject to the Bloch
condition\index{Bloch condition} \be\label{10.1}
\psi (x+L) = e^{ikL} \psi (x) \, ,
\ee
where $k$ is real and
denotes the crystal momentum. As a result, the spectrum shows energy
bands whose edges correspond to $kL = 0, \pi$, that is the wave function at
the band edges satisfy $\psi (x+L) = \pm \psi (x)$. For periodic potentials,
the band edge energies and wave functions are often called eigenvalues and
eigenfunctions, and we will use this terminology in these lectures. 
In particular
the ground state eigenvalue and eigenfunction refers to the bottom edge of the
lowest energy band.

Let us first
discuss the question of SUSY breaking for periodic potentials. Since
$H_1 = A^{\dag}A$ and $H_2 = AA^{\dag}$ are formally positive operators
 their spectrum is nonnegative and almost the
same. The caveat ``almost'' is
needed because the mapping between the positive energy states of the two does
not apply to zero energy states.

The Schr\"odinger equation for $H_{1,2}$ has zero energy modes given by
\be\label{10.3}
\psi_{0}^{(1,2)} (x) = \exp
\bigg (\mp \int^{x} dy W(y) \bigg ) \, ,
\ee
provided $\psi_{0}^{(1,2)}$ belong to the Hilbert space.
Supersymmetry is unbroken if at least one of the $\psi_{0}^{(1,2)}$
is a true
zero mode while otherwise it is dynamically broken. 

For a non-periodic potential we have seen that at most one
of the functions $\psi_{0}^{(1,2)}$ can be normalizable and hence an acceptable
eigenfunction. By convention we are choosing $W$ such that only $H_1$
(if at all) has a zero mode.

Let us now consider the case when $W$ (and hence $V_{1,2}$) are periodic
with period $L$. Now the eigenfunctions including the ground state
wave function must satisfy the Bloch condition 
(\ref{10.1}). But, in view of eq. (\ref{10.3}) we have
\be\label{10.4}
\psi_{0}^{(1,2)} (x+L) = e^{\pm \phi_L} \psi_{0}^{(1,2)} (x) \, ,
\ee
where
\be\label{10.5}
\phi_L = \int^{x+L}_{x} W(y) dy \, .
\ee
On comparing eqs. (\ref{10.1}) and (\ref{10.4}) it is clear that
for either of the wave functions $\psi_{0}^{(1,2)}$
to belong to the Hilbert
space, we must identify $\pm\phi_L=ikL$. But $\phi_L$ is real (since $W$ and
hence $V_{1,2}$ are assumed to be real), which means
that $\phi_L=0$. Thus, the two functions $\psi_{0}^{(1,2)}$ either
{\it both} belong to the Hilbert space, in which case they are strictly
periodic with period $L$: $\psi_{0}^{(1,2)}(x+L)=\psi_{0}^{(1,2)}(x)$,
or (when $\phi_L\neq 0$) {\em neither of them} belongs to the Hilbert space.
Thus in the periodic case, irrespective of whether SUSY is broken or
unbroken, the spectra of $V_{1,2}$ is always strictly isospectral.

To summarize, we see that
\be\label{10.6}
\phi_L = \int_0^{L} W(y) dy = 0
\ee
is a necessary condition for unbroken SUSY, and when
this condition is satisfied then $H_{1,2}$  have identical spectra, including
zero modes. In this case, using the known
eigenfunctions $\psi^{(1)}_n (x)$ of $V_{1}(x)$
one can
immediately
write down the corresponding (un-normalized) eigenfunctions $\psi^{(2)}_n (x)$
of $V_{2} (x)$. In particular, from eq. (7.3) the
ground state of $V_{2} (x)$ is given by
\be\label{10.6a}
\psi^{(2)}_{0} (x) = {1\over \psi^{(1)}_{0} (x)}= e^{\int^x W(y) ~dy }\,~ ,
\ee
while the excited states $\psi^{(2)}_{n} (x)$ are
obtained from $\psi^{(1)}_{n} (x)$
by using the relation
\be\label{10.6b}
\psi^{(2)}_{n} (x)
= [\frac{d}{dx} +W(x)] \psi^{(1)}_{n} (x) ~,~ (n \ge 1)~.
\ee
Thus by starting from an exactly solvable periodic potential $V_{1}(x)$,
one gets another  strictly isospectral periodic
potential $V_{2}(x)$. 

At this stage, it is worth pointing out that there are some special
classes of periodic superpotentials which trivially satisfy the
condition (\ref{10.6}) and hence for them SUSY is unbroken.
For example, suppose the superpotential is antisymmetric
on a half-period:
\be\label{10.7}
W(x+{L\over 2}) = - W(x) \, .
\ee
Then,
\be\label{10.8}
V_{1,2}(x+{L\over 2}) \equiv W^2 (x+{L\over 2}) \mp W(x+{L\over 2})
=V_{2,1} (x) \, .
\ee
Thus in this case $V_{1,2}$ are simply translations of one another by half a
period, and hence are essentially identical in shape. Therefore, they must
support exactly the same spectrum, as SUSY indeed tells us they do. Such a
pair of isospectral $V_{1,2}$ that are identical in shape are termed as
``self-isospectral''. A simple example of a
superpotential of this type is $W(x)={\rm cos} x$, so that
$V_{2}(x) = {\rm cos}^2 x - {\rm sin} x = V_{1} (x+\pi)$.
In a way, self-isospectral potentials are uninteresting since
in this case, SUSY will give us nothing new.

More generally, if a pair of periodic partner potentials $V_{1,2}$ are such
that $V_{2}$ is just the partner potential $V_{1}$
up to a discrete transformation-a translation by any constant amount, a
reflection, or both, then such a pair of partner potentials 
are termed as ``self-isospectral ''.
For example,
consider periodic superpotentials that are even functions of $x$:
\be\label{10.9}
W(-x) =  W(x) \, ,
\ee
but which also satisfy the condition (\ref{10.6}).
Since the function $dW(x)/dx$ is now odd hence it follows that
\be\label{10.10}
V_{1,2}(-x) = V_{2,1}(x) \, .
\ee
The partner potentials are then simply reflections of one another.
They therefore have the same shape and hence give rise to exactly the same
spectrum.
A simple example of a superpotential of this type is again
$W(x)={\rm cos} x$, so that
$V_{2}(x) = {\rm cos}^2 x - {\rm sin} x = V_{1}(-x)$.

It must be made clear here that not all periodic partner potentials are
self-isospectral even though they are strictly isospectral.
Consider for example, periodic superpotentials that are odd functions of $x$:
\be\label{10.11}
W(-x) = - W(x) \, .
\ee
Then the condition (\ref{10.6}) is satisfied trivially and hence SUSY is
unbroken. The function
$dW(x)/dx$ is even and thus $V_{1,2}(x)$ are also even. In this case,
$V_\pm(x)$ are not necessarily related by simple translations or reflections.
For example, the
superpotential $W(x)= A~{\rm sin}x + B~{\rm sin} 2x $ gives rise to an
isospectral pair which is not self-isospectral.
On the other hand, $W(x)= A~{\rm sin}x + B~{\rm sin} 3x $
gives rise to a self-isospectral pair since this $W$ satisfies the condition
(\ref{10.7}).

\subsection{Lam\'e Potentials and Their Supersymmetric Partners}

The classic text book example of a periodic potential which is often used to
demonstrate band structure is the Kronig-Penney model  \index{Kronig-Penney
model}, \be\label{10.12}
V(x) = \sum^{\infty}_{-\infty} V_{0} \delta (x-nL) \, .
\ee
It should be noted that the band edges of this model can only be computed by
solving a transcendental equation.

Another well studied class of periodic problems consists of the
Lam\'{e} potentials
\index{potentials!Lam\'{e}}
\be\label{10.13}
V(x,m) = p~m~ {\rm sn}^2 (x,m) \, , \ \ p \equiv a(a+1) \, .
\ee
Here ${\rm sn}(x,m)$ is a Jacobi elliptic function \index{Jacobi elliptic
function}  of real elliptic modulus
parameter $m (0 \le m \le 1)$ with period $4K(m)$, where $K(m)$ is the `` real
elliptic quarter period '' given by
\be\label{10.14}
K(m) = \int^{\frac{\pi}{2}}_{0} \frac{d\theta}{\sqrt{1-m\sin^2 \theta}} \, .
\ee
For simplicity, from now on, we will not explicitly display the modulus
parameter $m$ as an argument of Jacobi elliptic functions unless necessary.
Note that the
elliptic function potentials (\ref{10.13}) have period $2K(m)$. They
will be referred to as Lam\'e potentials,
since the corresponding Schr\"odinger
equation is called the Lam\'e equation
in the mathematics literature. It is known that for
any integer value $a = 1,2,3,...$,
the corresponding Lam\'{e} potential has $a$ bound bands followed
by a continuum band. All the band edge energies and the corresponding
wave functions are analytically
known. We shall now apply the formalism of SUSY QM and calculate the SUSY
partner potentials corresponding to the Lam\'{e} potentials as given by
eq. (\ref{10.13}) and show that even though $a=1$ Lam\'{e} partners are
self-isospectral, for $a \ge 2$ they are not self-isospectral. Consequently, 
SUSY QM generates new exactly solvable periodic problems!

Before we start our discussion, it is worth mentioning a few basic properties
of the Jacobi elliptic functions \index{Jacobi elliptic function} $\sn  \,  x,
\cn  \,  x$ and $\dn  \,  x$ which we shall be using in this discussion.
First of all, whereas $\sn  \,  x$ and $\cn  \,  x$ have period $4K(m)$,
$\dn  \,  x$ has period $2K(m)$
[i.e. $\dn(x+2K(m)) = \dn  \,  x$]. They are related to
each other by
\be\label{10.15}
m~{\rm sn}^2 x = m -m~{\rm cn}^2 x = 1 - {\rm dn}^2 x \, .
\ee
Further,
\be\label{10.16}
\frac{d}{dx} \sn  \,  x = \cn  \,  x ~\dn  \,  x \, ; 
 \frac{d}{dx} \cn  \,  x = -\sn  \,  x ~\dn  \,  x \, , 
\frac{d}{dx} \dn  \,  x = -m~ \sn  \,  x ~\cn  \,  x \, .
\ee
Besides
\be\label{10.16a}
\sn(x+K) = \frac{\cn  \,  x}{\dn  \,  x} \, ; \
~\cn(x+K) = -\sqrt{1-m}\frac{\sn  \,  x}{\dn  \,  x} \, ; \
~\dn(x+K) = \frac{\sqrt{1-m}}{\dn  \,  x} \, .
\ee
Finally, for $m=1 (0)$, these functions reduce to the familiar hyperbolic
(trigonometric) functions, i.e.
\bea\label{10.17}
&& \sn(x,m=1) = \tanh x \, ;~ \cn(x,m=1) = \sech ~x \, ;
~\dn(x,m=1) = \sech ~x \, , \nonumber \\
&& \sn(x,m=0) = \sin x \, ; ~\cn(x,m=0) = \cos x \, ;
~\dn(x,m=0) = 1 \, .
\eea

It may be noted that when $m=1$, the Lam\'e potentials (\ref{10.13}) reduce to
the well known P\"oschl-Teller potentials 
\be\label{10.18}
V(x,m=1) = a(a+1) - a(a+1) \sech^2 x \, ,
\ee
which for integer $a$ are reflectionless and have $a$ bound
states. It is worth adding here that in the limit $m \rightarrow 1$, $K(m)$
tends to $\infty$ and the periodic nature of the potential is obscure.
On the other hand, when $m=0$, the Lam\'e potential (\ref{10.13})
vanishes and one has a rigid rotator 
problem (of period $2K(m=0) = \pi$),
whose energy eigenvalues are at $E = 0,1,4,9,...$ with all the nonzero energy
eigenvalues being two-fold degenerate.

Finally, it may be noted that the Schr\"odinger equation for finding the
eigenstates for an arbitrary periodic potential is called Hill's equation
in the mathematical literature.
A general property of the Hill's equation is the
oscillation theorem
which states
that for a potential with period $L$, the band edge wave functions
arranged in order of increasing energy
$E_0 \le E_1 \le E_2 \le E_3 \le E_4 \le E_5 \le E_6 \le ...$
are of period $L,2L,2L,L,L,2L,2L,...$~. The corresponding number of (wave
function) nodes in the interval $L$ are $0,1,1,2,2,3,3,...$ and the energy
band gaps are given by $\Delta_1 \equiv E_2 - E_1,~\Delta_2 \equiv E_4 - E_3,~
\Delta_3 \equiv E_6 - E_5,~... $~. We shall see that the expected $m=0$ limit
and the oscillation theorem are very useful in making sure that all band
edge eigenstates have been properly determined or if some have been missed.

Let us first consider the Lam\'e potential (\ref{10.13}) with $a=1$ and show
that in this case the SUSY partner potentials are self-isospectral.
The Schr\"odinger equation for the Lam\'e potential with $a=1$ can be solved
exactly and it is well known that in this case the spectrum consists of a
single bound band followed by a continuum band. In particular, the eigenstates for the
lower and upper edge of the bound band are given by
\be\label{10.19}
E_0 = m \, ; \ \ \psi_{0} (x) = \dn  \,  x \, ,
\ee
\be\label{10.20}
E_1 = 1 \, ; \ \ \psi_{1} (x) = \cn  \,  x \, .
\ee
On the other hand, the eigenstate for the lower edge of the continuum band is
given by
\be\label{10.21}
E_2 = 1+m \, ; \ \ \psi_{2} (x) = \sn  \,  x \, .
\ee
i.e. it extends from $1+m$ to $\infty$. Note that
at $m=0$ the energy eigenvalues are at $0,1$ as expected for a rigid rotator
and as $m \rightarrow 1$, one gets $V(x) \rightarrow 2 - 2 ~\sech^2 x$,
the band width $1-m$ vanishes as expected, and one has an energy level at $E=1$.

Using eq. (\ref{10.19}) the corresponding superpotential turns out to be
\be\label{10.22}
W(x) = m \frac{\sn  \,  x ~\cn  \,  x}{\dn  \,  x} \, ,
\ee
On making use of eq. (\ref{10.16a}) it is easily shown 
that this $W$ satisfies the condition (\ref{10.7})
and hence
the corresponding partner potentials are indeed self-isospectral.

The Lam\'{e} potential (\ref{10.13}) with $a=1$,
is one of the rare periodic potentials for
which the dispersion relation\index{dispersion relation}
between $E$ and crystal momentum $k$ is known in
a closed form. 

In view of this result for $a=1$, one might think that even
for higher integer
values of $a$, the two partner potentials would be self-isospectral. However,
this is not so, and in fact for any integer $a\, (\ge 2)$, we obtain  new exactly
solvable periodic potential.
As an illustration, consider the Lam\'{e} potential (\ref{10.13}) with $a=2$.
For the $a$ = 2 case,
the Lam\'{e} potential has 2 bound bands and a
continuum band. The energies and wave functions of the
five band edges are well known. The lowest energy band
ranges from $2+2m-2\delta$ to $1+m$, the second energy band ranges
from $1+4m$ to $4+m$ and the continuum starts at energy $2+2m+2\delta$,
where $\delta = \sqrt{1-m+m^2}$. 

 Note that in the
interval $2K(m)$ corresponding to the
period of the  Lam\'{e} potential, the number of nodes increase with energy.
In order to use the SUSY QM formalism,
we must shift the Lam\'{e} potential by a constant to ensure
that the ground state (i.e. the lower edge of the lowest band) has energy
$E = 0$. As a result, the potential
\be\label{10.23}
V_{1} (x) = -2-2m+2\delta+6m \,{\rm sn}^2 x \, ,
\ee
has its ground state energy at zero with the  corresponding un-normalized
wave function
\be\label{10.24}
\psi^{(1)}_{0} (x) = 1+m+\delta-3m \,{\rm sn}^2 x~.
\ee
The corresponding superpotential is
\be\label{10.25}
W = - {d\over dx} \log \psi^{(1)}_{0} (x)
= {6m \sn  \,  x ~~ \cn  \,  x \,\dn  \,  x\over
\psi^{(1)}_{0} (x)} \, ,
\ee
and hence the partner potential  corresponding to
(\ref{10.23}) is
\be\label{10.26}
V_{2}(x) = -V_{1}(x)+
{72 m^2\, {\rm sn}^2 x \,{\rm cn}^2 x \,{\rm dn}^2 x\over [1+m+\delta-3m \,{\rm
sn}^2 x]^2}~~. \ee

Although the SUSY QM formalism guarantees that the potentials $V_{1,2}$
are isospectral, they are not
self-isospectral, since they do not satisfy
eq. (\ref{10.8}).
Therefore, $V_{2}(x)$ as given by eq. (\ref{10.26}) is a new
periodic potential which is strictly isospectral
to the potential (\ref{10.23}) and hence it also has 2 bound bands and a
continuum band. Similar conclusions are also valid for Lame potentials
(\ref{10.13}) with $a=3,4,5,...$.

\end{document}